\documentclass[11pt,a4paper,twoside]{article}

\usepackage{amsmath,amsthm,amstext,amscd,amssymb,euscript,mathrsfs}
\usepackage{calrsfs}
\usepackage{epsfig}

\newcommand{\R}{\mathbb R}
\newcommand{\N}{\mathbb N}

\newcommand{\E}{\mathbb E}

\newcommand{\epsi}{\ensuremath{\epsilon}}

\newcommand{\pee}{\ensuremath{\mathbb{P}}}

\def\1{{\mathchoice {\rm 1\mskip-4mu l} {\rm 1\mskip-4mu l}
{\rm 1\mskip-4.5mu l} {\rm 1\mskip-5mu l}}}

\newtheorem{theorem}{{\small T}{\scriptsize HEOREM}}[section]
\newtheorem{corollary}{{\bf{\small C}{\scriptsize OROLLARY}}}[section]
\newtheorem{proposition}{{\bf{\small P}{\scriptsize ROPOSITION}}}[section]
\newtheorem{lemma}{{\bf{\small L}{\scriptsize EMMA}}}[section]
\newtheorem{remark}{{\bf{\small R}{\scriptsize EMARK}}}[section]
\newtheorem{definition}{{\bf{\small D}{\scriptsize EFINITION}}}[section]

\renewenvironment{proof}[1]
{\noindent{{\bf{\small{ P}{\scriptsize ROOF}}}.}\hspace{0.1cm} #1} {$\;\qed$\newline}

\newcommand{\beq}{\begin{eqnarray}}
\newcommand{\eeq}{\end{eqnarray}}

\newcommand{\ba}{\begin{align*}}
\newcommand{\ea}{\end{align*}}

\newcommand{\be}{\begin{equation}}
\newcommand{\ee}{\end{equation}}

\newcommand{\bl}{\begin{lemma}}
\newcommand{\el}{\end{lemma}}

\newcommand{\br}{\begin{remark}}
\newcommand{\er}{\end{remark}}

\newcommand{\bt}{\begin{theorem}}
\newcommand{\et}{\end{theorem}}

\newcommand{\bd}{\begin{definition}}
\newcommand{\ed}{\end{definition}}

\newcommand{\bp}{\begin{proposition}}
\newcommand{\ep}{\end{proposition}}

\newcommand{\bc}{\begin{corollary}}
\newcommand{\ec}{\end{corollary}}

\newcommand{\bpr}{\begin{proof}}
\newcommand{\epr}{\end{proof}}

\newcommand{\bi}{\begin{itemize}}
\newcommand{\ei}{\end{itemize}}

\newcommand{\ben}{\begin{enumerate}}
\newcommand{\een}{\end{enumerate}}

\newcommand{\qwe}{{\ldots}}

\begin{document}
\title{Weak coupling limits in a stochastic model of heat conduction}  
\author{ 
Frank Redig$^{\textup{{\tiny(a)}}}$, Kiamars Vafayi$^{\textup{{\tiny(b)}}}$\\
{\small $^{\textup{a)}}$ IMAPP, Radboud Universiteit Nijmegen}\\
{\small Heyendaalse weg 135,
6525 AJ Nijmegen, The Netherlands}\\
{\small f.redig@math.ru.nl}\\
{\small $^{\textup{(b)}}$ Mathematisch Instituut Universiteit Leiden}\\
{\small Niels Bohrweg 1, 2333 CA Leiden, The Netherlands}\\
{\small vafayi@math.leidenuniv.nl}\\
} 

\maketitle

\begin{abstract}
We study the Brownian momentum process, a model of heat conduction, weakly coupled to
heat baths. 
In two different settings of weak coupling to the heat baths,
we study the non-equilibrium steady state
and its proximity to the local equilibrium measure in terms
of the strength of coupling.
For three and four site systems, we obtain the two-point correlation function
and show it is generically not multilinear.
\bigskip

\noindent
{\bf Keywords}: weak coupling limit, local equilibrium, 
Brownian momentum process, inclusion process, duality.

\end{abstract}

%%%%%%%%%%%%%%%%%%%%%%%%%%%%%%%%%%%%%%%%%%%%%%%%%%%%%
\section{Introduction}
In the study of non-equilibrium systems, exactly solvable models can
serve as test-cases with which general statements about non-equilibrium, such as
in \cite{bg09}, \cite{mn} 
can be tested. Recently, in \cite{gkr}, \cite{gkrv}, \cite{grv},
we studied the Brownian momentum process (BMP) and showed that this models is exactly solvable
via duality with a particle system,
the symmetric
inclusion process. In this paper, we 
look at the close-to-equilibrium states of the BMP. 
First, we consider a close-to-equilibrium scenario where the
temperature of the right heat bath is close to the temperature of the left heat bath, 
and show that the distance between the local equilibrium measure and the
true non-equilibrium steady state is of order at most the square of the temperature
difference, in agreement with the theory of Mc Lennan ensembles, see \cite{mn}. Next, we consider a situation
where the linear chain is coupled weakly to heat baths to left and right ends (with
fixed and different temperatures), and study which
equilibrium measure is selected in the limit where the coupling strength $\lambda$ tends
to zero, as well as how far the true non-equilibrium steady state is from
the local equilibrium measure for small coupling strengths. The temperature
profile can be computed for all values of $\lambda$ and is only
linear in the chain including the extra sites associated to the heat baths
for $\lambda=1$, and linear if these sites are not included for
all values of $\lambda>0$.
Finally, we explicitly compute the two-point correlation for all $\lambda>0$ for a
three and four sites system and show that the multilinear ansatz of the
two-point function introduced in \cite{gkr}, see also \cite{bg09}, \cite{derr} fails for a system of four sites, except when $\lambda=1$.
\section{The model}
The Brownian momentum process on a linear chain $\{1,\ldots,N\}$ coupled
at the left and right end to a heat bath is a Markov process $\{x(t):t\geq 0\}$ on
the state space $\Omega_N=\R^{\{1,\ldots,N\}}$. The configuration
$x(t)= x_i(t):i\in\{1,\ldots N\}$ is interpreted as momenta associated
to the sites 
$i\in \{1,\ldots,N\}$.
The process is defined via its generator
working on the core of smooth functions
$f:\Omega_N \to\R$ which is given by
\be\label{bmpgen}
L=\lambda B_{1}+\lambda B_{N}+\sum_{i,j}^{N}p(i,j)L_{i,j}
\ee
with
\[
L_{i,j}=\left(x_{i}\partial_{j}-x_{j}\partial_{i}\right)^{2} 
\]
and where $\partial_j$ is shorthand for $\frac{\partial}{\partial x_j}$.
The underlying random walk transition rate $p(i,j)$
is chosen to be symmetric and
nearest neighbor, i.e., $p_{i,i+1}=p_{i+1,i}=1, i\in \{1,\ldots,N-1\}$, $p(i,j)=0$ otherwise.
% $p(i,j)=p(j,i)= p(j-i) \geq 0$ are rates of an irreducible
%symmetric and translation invariant continuous-time random walk on $\{1,\ldots,N\}$.
Since $L_{i,j}=L_{j,i}$ the symmetry of $p(i,j)$ is no loss of generality.

The boundary operators $B_1, B_N$ model the contact with the heat baths, and are
chosen to be Ornstein-Uhlenbeck generators corresponding to the temperatures of
the left and right heat bath, i.e.,
\[
B_{1}=T_{L}\partial_{1}^{2}-x_{1}\partial_{1}\]

\[
B_{N}=T_{R}\partial_{N}^{2}-x_{N}\partial_{N}\]

Finally, $\lambda>0$ measures the strength of the coupling to the heat baths.
The process with generator
\eqref{bmpgen} is abbreviated as $BMP_\lambda$.

If $T_L=T_R=T$, then, for all $\lambda>0$, the unique stationary measure of the process
$\{x(t):t\geq 0\}$ is the product of Gaussian measures with mean zero
and variance $T$. If $T_L\not= T_R$ there exists a unique stationary measure; the so-called non-equilibrium steady state denoted by $\mu^\lambda_{T_L,T_R}$.
The existence and uniqueness of the measure $\mu^\lambda_{T_L,T_R}$ follows
from duality (see next section).

We will look at two different close-to-equilibrium scenarios:
\ben
\item $\lambda =1$, $T_R= T_L+ \epsi$ and $\epsi\to 0$,
\item $T_L\not= T_R$, and $\lambda\to 0$.
\een
In both cases we look at the behavior of the measure $\mu^\lambda_{T_L,T_R}$, in case
two, as $\lambda\to 0$, and in case one as $\epsi\to 0$. Since for $\lambda=0$, the
system has infinitely many equilibrium measures, in the second case it
is of interest to find out which of these measure is selected in the limit $\lambda\to 0$.
Both in the first and second case, we want to understand how close the true non-equilibrium steady state
is to the local equilibrium measure.
\section{Duality}
The $BMP_\lambda$ can be analyzed via duality. The dual process is
an interacting particle system, the so-called symmetric inclusion process
\cite{grv}, where particles are jumping on the lattice $\{0,1,\ldots, N, N+1\}$ 
and interacting by ``inclusion'' (i.e., particles at site $i$ can attract
particles at site $j$).
The ``extra sites'' $0, N+1$ -associated to the heat baths- are absorbing.
I.e., a dual particle configuration is a map
\[
 \xi: \{0,\ldots, N+1\}\to \N
\]
specifying at each site the number of particles present at that site. The
space of dual particle configurations is denoted by $\Omega^d_N$
For $\xi \in \Omega^d_N$, $\xi^{i,j}$ denotes the configuration obtained from
$\xi$ by removing a particle from $i$ and putting it at $j$.

The generator of the dual process then reads
\begin{eqnarray}\label{sipgen}
&&L_{d}\phi(\xi)=2\lambda\xi_{1}[\phi(\xi^{1,0})-\phi(\xi)]+\nonumber\\
&+&\sum_{i,j=1}^{N-1}p(i,j)\left(2\xi_{j}(2\xi_{i}+1)[\phi(\xi^{j,i})-\phi(\xi)]+
2\xi_{i}(2\xi_{j}+1)[\phi(\xi^{i,j})-\phi(\xi)]\right)\nonumber\\
&&+2\lambda\xi_{N}[\phi(\xi^{N,N+1})-\phi(\xi)]
\end{eqnarray}
In words, this means particles
at site $i$ jump to $j$ at rate $2p(i,j)(2\xi_{j}+1)$.
At the boundary site $1$ (resp.\ $N$) particles can jump at rate $2\lambda$ to the site $0$ (resp.\ $N+1$) where they are absorbed. 
Absorbed particles do not interact with non-absorbed ones. 
The dual process is abbreviated as $SIP_\lambda$.
The duality functions for duality between $BMP_\lambda$ and $SIP_\lambda$
are independent of $\lambda$ and given by
\[
D(\xi, x)=T_{L}^{\xi_{0}}T_{R}^{\xi_{N+1}}\prod_{i=1}^{N}\frac{x_{i}^{2\xi_{i}}}{(2\xi_{i}-1)!!}\]
for $\xi \in \Omega_N^d$ a dual particle configuration, and $x\in \Omega_N$.

The duality relation then reads
\be\label{dualrel}
LD(\xi, x)=L_{d}D(\xi, x)
\ee
where $L$ works on $x$ and $L_d$ on $\xi$.
By passing to the semigroup, from \eqref{dualrel} we obtain
the duality relation 
\be\label{dualrell}
 \E_x D(\xi,x(t))= \E^d_\xi D(\xi(t),x)
\ee
where $\E_x$ is expectation in $BMP_\lambda$
starting from $x\in\Omega_N$, and $\E^d_\xi$ is expectation in $SIP_\lambda$ starting from $\xi\in\Omega^d_N$.

For $\xi\in \Omega^d_N$ we denote $|\xi|=\sum_{i=0}^{N+1} \xi_i$ the
total number of particles in $\xi$.
Since eventually all particles in a particle configuration $\xi\in \Omega^d_N$ will
be absorbed, we have a unique stationary distribution $\mu_{T_L,T_R}^{\lambda}$ with
\be\label{statmeas}
 \int D(\xi, x) \mu_{T_L,T_R}^{\lambda} (dx) = \sum_{k,l: k+l= |\xi|} T_L^k T_R^l \pee^d_\xi \left(\xi(t=\infty)= k\delta_0 +l \delta_{N+1}\right) 
\ee
where $\xi(t=\infty)$ denotes the final configuration when all particles
are absorbed and $k\delta_0 +l \delta_{N+1}$ the configuration
with $k$ particles at $0$ and $l$ particles at $N+1$.

\section{Temperature profile}
The local temperature at site $i\in \{1,\ldots,N\}$ is defined as
\[
 T_i = \int x_i^2 \mu^\lambda_{T_L,T_R} (dx)
\]
and by definition $T_0= T_L, T_{N+1}=T_R$.
We say that the temperature profile is linear in the lattice interval $[K,L]$ if
there exist $a,b\in \R$ with $T_i = ai +b$, for all $i\in [K,L]$.
For the computation of the temperature profile we only need a single dual walker,
which performs a continuous-time random walk with rates $2p(i,j)$ and
absorption at rate $2\lambda$ from the sites $1,N$.

Indeed, using \eqref{statmeas}
we have
\be\label{tempran}
 T_i = T_L \pee^d_{\delta_i} \left(\xi(\infty)= \delta_0\right) + 
T_R \left(1- \pee^d_{\delta_i} \left(\xi(\infty)=\delta_0\right)\right)
\ee

From this expression, one obtains the following equations for
the temperature profile:
\beq\label{tempeq}
\sum_{i=1}^N p(i,1) T_i &=& T_1-\lambda(T_L-T_1) 
\nonumber\\
\sum_{i=1}^N p(i,k) T_i &=& T_k
\nonumber\\
\sum_{i=1}^N p(i,N) T_i &=& T_N-\lambda (T_R-T_N)
\eeq
The second equation expresses that the temperature profile is
a harmonic function of the transition probabilities, whereas
the first and third equation are boundary conditions.
In the case $\lambda=1$ and $p$ corresponding to the simple nearest neighbor random walk,
the equation for $T_i,  i=0,\ldots,N$ is the discrete Laplace equation, which
gives a linear temperature profile in $[0,N+1]$.

\br
In this paper we restrict to the symmetric
nearest neighbor walk kernel $p(i,j)$.
The equations \eqref{tempeq} hold for general symmetric $p(i,j)$. However, in the cases where it is not translation-invariant and/or
not nearest neighbor, the temperature profile will not be linear.
\er
We have the following theorem that follows immediately from
the equations \eqref{tempeq}. 
\bt\label{tempthm}
For all $\lambda >0$, the temperature profile
is linear in $[1,N]$ and
is given by
\be\label{nntemp}
T_i= a i + b
\ee
$i=1,\qwe, N$ with
\[
a=\frac{\lambda(T_{R}-T_{L})}{\lambda(N-1)+2}\]
\[
b=\frac{T_{L}+T_{R}+\lambda(NT_{L}-T_{R})}{\lambda(N-1)+2}\]
\et
We can now look at different limiting cases:
\ben
\item In the case $\lambda=1$ we recover the result from
\cite{gkr}:
\[
T=T_{L}+\frac{T_{R}-T_{L}}{N+1}i\]
In this case (only) the temperature profile is linear in $[0,N+1]$.
\item In the limit $\lambda\to 0$ we obtain for all $i \in \{1,\ldots,N\}$
\[
 \lim_{\lambda\to 0}T^{(\lambda)}_i= \frac{T_L+T_R}{2} 
\]
\item In the limit $\lambda\to\infty$ we obtain
$T_1=T_L, T_{N}=T_R$ and the profile is linear
in $[1,N]$, similar to a system
with $\lambda=1$ and $N-2$ sites.
\item In the limit $N\to\infty$, such that $i/N\to r\in [0,1]$ fixed,
\[
 \lim_{N\to\infty, \frac{i}{N}\to r} T_i = T_L + r(T_R-T_L)
\]
This means that the macroscopic profile
is linear and
does not depend on $\lambda$.
\een

\br
The expectation of the heat current in the steady state in the system is $J=T_{i+1}-T_{i}$. Heat conductivity $\kappa$ is defined via the equation $J=\kappa \Delta T$.
From Theorem \ref{tempthm} it follows that $\kappa=\frac{\lambda}{\lambda(N-1)+2}$ which is independent of the temperature (i.e. the system obeys the Fourier's law
for all values of $\lambda>0$).
\er

\section{The stationary measure for $\epsi\to 0$}
We consider the first weak coupling setting, i.e, $\lambda=1$, $T_R= T_L+\epsi$.
We will prove that up to corrections of order $\epsi^2$, the stationary
measure is given by a product of Gaussian measures 
corresponding to the temperature profile, i.e., the local equilibrium
measure.

Let us denote
this local equilibrium measure
\[
 \nu_{T_L, T_R} =\otimes_{i=1}^N G_{T_i} (x_i) dx_i
\]
with $T_i$ given by \eqref{nntemp},
\[
 G_T (x) =\frac{1}{\sqrt{2\pi T}} \exp (-x^2/2T)
\]
and $\mu_{T_L,T_L+\epsi}$ the true non-equilibrium steady state (with $\lambda=1)$.
Then we have the following result.
\bt
The true equilibrium measure and the local equilibrium measure are at most
order $\epsi^2$ apart, i.e., there exists
$\epsi_0>0$ such that for all $\xi\in\Omega^d_N$ there exists a constant $C=C(\xi) <\infty$ such that for all $0\leq \epsi\leq \epsi_0$ we have
\be\label{colibri}
\left|\int D(\xi, x) \mu_{T_L,T_L+\epsi} (dx) -\int D(\xi, x) \nu_{T_L,T_L+\epsi} (dx)\right | \leq C(\xi) \epsi^2
\ee
\et
\bpr
For the local equilibrium measure we have
\be\label{ping}
 \int D(\xi, x) \nu_{T_L,T_L+\epsi} (dx)= \prod_{i=1}^N T_i^{\xi_i}
\ee
expanding this up to order $\epsilon$ we find,
\[
\prod_{i}T_{i}^{\xi_{i}}=\prod_{i}\left(T_{L}+\frac{\epsilon i}{N+1}\right)^{\xi_{i}}=T_{L}^{|\xi|}\left(1+\frac{\epsilon}{T_{L}(N+1)}\sum_{i}i\xi_{i}\right) + O(\epsi^2)
\]
Start now from \eqref{statmeas} and expand up to order $\epsi$:
\beq\label{pong}
&&\int D(\xi,x)\mu_{T_{L}T_L +\epsi}(dx)
\nonumber\\
&=&T_{L}^{|\xi|}\left(1+\frac{\epsilon}{T_{L}}
\sum_{
k,l:
k+l=|\xi|}l \pee^d_\xi(\xi(\infty)=k\delta_{0}+l\delta_{N+1})\right) + O (\epsi^2)
\eeq
Upon identification of \eqref{ping} and \eqref{pong} we see that we have to prove
\beq
\sum_{k,l:
k+l=|\xi|} l\pee^d_\xi(\xi(\infty)=k\delta_{0}+l\delta_{N+1})
&=&\E_{\xi}(\xi_{\infty}(N+1))
\nonumber\\
&=&
\frac{1}{(N+1)}\sum_{i=0}^{N+1}i\xi_{i}=:\psi(\xi)
\eeq
The function $\phi(\xi):=\E_\xi (\xi_{\infty} (N+1))$ is the harmonic function
for the dual process, i.e.,
\[
 L_d \phi =0
\]
which satisfies the boundary conditions
\be\label{krawa}
 \phi\left(k\delta_0 +\sum_{i=1}^N \xi_i \delta_i + l\delta_{N+1}\right) = 
\phi\left(\sum_{i=1}^N \xi_i \delta_i\right) + l
\ee
Therefore, it suffices to show that 
\[
 \frac{1}{(N+1)}\sum_{i=0}^{N+1}i\xi_{i}=:\psi(\xi)
\]
both satisfies 
\[
 L_d \psi=0
\]
and the boundary conditions \eqref{krawa}. That $\psi$ satisfies the boundary conditions
is immediately clear. 
The fact that $\psi$ is harmonic follows from explicit computation:
\begin{eqnarray*}
L_{d}\psi(\xi)&=&2\xi_{1}[\psi(\xi^{1,0})-\psi(\xi)]\\
&+& \sum_{i=1}^{N-1}\left(2\xi_{i+1}(2\xi_{i}+1)[\psi(\xi^{i+1,i})-\psi(\xi)]+2\xi_{i}(2\xi_{i+1}+1)[\psi(\xi^{i,i+1})-\psi(\xi)]\right)\\
&+&2\xi_{N}[\psi(\xi^{N,N+1})-\psi(\xi)]
\\
&=&\frac{1}{N+1}\Big(2\xi_{1}[-1]+\\
&+&\sum_{i=1}^{N-1}\left(2\xi_{i+1}(2\xi_{i}+1)[-1]+2\xi_{i}(2\xi_{i+1}+1)[+1]\right)\\
&+& 2\xi_{N}[+1]\Big)
\\
&=&\frac{1}{N+1}\left(2\xi_{1}[-1]+
+2\sum_{i=1}^{N-1}\left(\xi_{i}-\xi_{i+1}\right)
+2\xi_{N}[+1]\right)
\end{eqnarray*}

and since $\sum_{i=1}^{N-1}\left(\xi_{i}-\xi_{i+1}\right)=\xi_{1}-\xi_{N}$
we indeed have \[
L_{d}\psi(\xi)=0\]
\epr
\section{The case $\lambda\to 0$}

Next, we consider the second weak coupling setting, i.e.,
we fix $T_L\not=T_R$ and study the behavior of the measure
$\mu^\lambda_{T_L,T_R}$ 
as a function of $\lambda$.

In this case, the local equilibrium measure is the product of Gaussian measures
corresponding to the temperature profile 
\eqref{nntemp}, i.e.,
we have to compare
$\mu^\lambda_{T_L,T_R}$ with $\nu^\lambda_{T_L,T_R}$ where
\[
 \nu^\lambda_{T_L,T_R} = \otimes_{i=1}^N G_{T^\lambda_i}(x_i) (dx_i)
\]
where 
$T^\lambda_i$ is given by \eqref{nntemp}.
Denote
\be\label{bobo}
 \phi (\xi) = \int D(\xi,x) \ \mu^\lambda_{T_L,T_R} (dx)
\ee
then $\phi$ is the harmonic function of the dual generator satisfying
the boundary conditions
\[
\phi(\xi^{*}=\xi+k\delta_{0}+l\delta_{N+1})=\psi(\xi).\psi(k\delta_{0}+l\delta_{N+1})=T_{L}^{k}T_{R}^{l}\psi(\xi)\]
On the other hand
if we put
\be\label{bibi}
 \psi (\xi) := \int D(\xi,x) \ \nu^\lambda_{T_L,T_R} (dx)=T_{L}^{k}T_{R}^{l}\prod_{i}(T_{i}^{(\lambda)})^{\xi_{i}}
\ee
then we see immediately that $\psi$ satisfies the boundary
conditions.

We will now first prove
\bl 
There exists $\lambda_0>0$ such that for all $\xi\in\Omega^d_N$ there exists $A(\xi)>0$ such that
for all $0<\lambda\leq \lambda_0$ we have
\[
 |\left(L_d\psi\right) (\xi) |\leq \lambda^2 A(\xi) 
\]
In particular, since there is only a finite number of dual particle configurations
with total number of particles equal to $K$, we have, for all $0<\lambda\leq \lambda_0$
\[
 \sup_{\xi:|\xi|=K}|\left(L_d\psi\right) (\xi) |\leq  C(K) \lambda^2
\]
for some $C(K)>0$
\el
\bpr
Compute

\begin{eqnarray*}
L_{d}\psi(\xi)&=&2\psi(\xi)\left(\lambda\xi_{1}\left(\frac{T_{L}}{T_{1}}-1\right)+\lambda\xi_{N}\left(\frac{T_{R}}{T_{N}}-1\right)
\right)
\\
 &+&2\psi(\xi)\left(
\sum_{i=1}^{N-1}\left(\xi_{i+1}(2\xi_{i}+1)\left(\frac{T_{i}}{T_{i+1}}-1\right)+\xi_{i}(2\xi_{i+1}+1)\left(\frac{T_{i+1}}{T_{i}}-1\right)\right)\right)
\end{eqnarray*}

Put
$T_{R}-T_{N}=T_{1}-T_{L}=:\gamma$

\begin{eqnarray}\label{bamboe}
&&L_{d}\psi(\xi)=2\psi(\xi)\left(\lambda\xi_{1}\left(\frac{-\gamma}{T_{1}}\right)+\lambda\xi_{N}\left(\frac{\gamma}{T_{N}}\right)\right)\nonumber\\
&+&2\psi(\xi)\left(\sum_{i=1}^{N-1}\left(2\xi_{i+1}\xi_{i}\frac{(T_{i}-T_{i+1})^{2}}{T_{i}T_{i+1}}+(T_{i}-T_{i+1})\left(\frac{\xi_{i+1}}{T_{i+1}}-\frac{\xi_{i}}{T_{i}}\right)\right)\right)
\end{eqnarray}

Remember from Theorem \ref{tempthm} that
$T_{i}=\lambda ai+b$,
hence
$T_{i}-T_{i+1}=-\lambda \alpha$, with
\[
\lambda \alpha=\frac{\lambda(T_{R}-T_{L})}{\lambda(N-1)+2}\]

\[
b=\frac{T_{L}+T_{R}+\lambda(NT_{L}-T_{R})}{\lambda(N-1)+2}\]

%\[
%\gamma=\frac{T_{L}+T_{R}+\lambda(NT_{L}-T_{R})}{\lambda(N-1)+2}+\frac{\lambda(T_{R}-T_{L})}%{\lambda(N-1)+2}-T_{L}\]

%\[
%\gamma=\frac{T_{L}+T_{R}+\lambda(NT_{L}-T_{R})+\lambda(T_{R}-T_{L})-T_{L}\lambda(N-1)-2T_{L}}%{\lambda(N-1)+2}\]

We find
\[
\gamma=\frac{T_{R}-T_{L}}{\lambda(N-1)+2}=\alpha\]

and hence, from \eqref{bamboe}
\[
L_{d}\psi(\xi)=
2\psi(\xi)\left(\lambda\xi_{1}
[\frac{-\alpha}{T_{1}}]+\lambda\xi_{N}[\frac{\alpha}{T_{N}}]+
\sum_{i=1}^{N-1}\left(2\lambda^{2}\alpha^2 \frac{\xi_{i}\xi_{i+1}}{T_i T_{i+1}}-\lambda \alpha\left(\frac{\xi_{i+1}}{T_{i+1}}-\frac{\xi_{i}}{T_{i}}\right)\right)\right)\]

We then see that the first order terms form a vanishing telescopic sum:

\begin{eqnarray*}
\sum_{i=1}^{N-1}\left(\frac{\xi_{i}}{T_{i}}-\frac{\xi_{i+1}}{T_{i+1}}\right)=\frac{\xi_{1}}{T_{1}}-\frac{\xi_{N}}{T_{N}}
\end{eqnarray*}

%\begin{eqnarray*}
%&&\xi_{1}[\frac{-\gamma}{T_{1}}]+\xi_{N}[\frac{\gamma}{T_{N}}]+a\sum_{i=1}^{N-1}\left(\frac{\xi_{i}}{T_{i}}-\frac{\xi_{i+1}}{T_{i+1}}\right)
%\\
%&=&\xi_{1}[\frac{-\gamma}{T_{1}}]+\xi_{N}[\frac{\gamma}{T_{N}}]+a\left(\frac{\xi_{1}}{T_{1}}-\frac{\xi_{N}}{T_{N}}\right)
%\\
%&=&\left(a-\gamma\right)\left(\frac{\xi_{1}}{T_{1}}-\frac{\xi_{N}}{T_{N}}\right)
%\\
%&=& 0
%\end{eqnarray*}

and therefore;

\[
L_{d}\psi(\xi)=4\lambda^{2}a^{2}\psi(\xi)\sum_{i=1}^{N-1}\left(\frac{\xi_{i+1}\xi_{i}}{T_{i}T_{i+1}}\right)\]

\epr
Given this result, we will prove that the measures $\nu^\lambda_{T_L,T_R}$ and
$\mu^\lambda_{T_L,T_R}$ are at most order $O(\lambda\log(1/\lambda))$ apart as $\lambda\to 0$.
\bt\label{lambdathm} 
Let $\phi, \psi$ be the functions defined in \eqref{bobo} and \eqref{bibi}, then we have the following.
There exists $\lambda_0>0$, such that for all $\xi\in\Omega^d_N$ there is $C(\xi)>0$, such that for all $0<\lambda\leq \lambda_0$  
\be\label{papegaai}
 |\phi(\xi) -\psi(\xi)|\leq C(\xi) \lambda \log\frac1{\lambda}
\ee
as a consequence,
\[
\lim_{\lambda\to 0}\mu^\lambda_{T_L,T_R} = \otimes_{i=1}^N G_{\frac{T_L+T_R}{2}}
\left(x_i\right)dx_i
\]
i.e.,
in the limit $\lambda\to 0$, the equibrium measure corresponding to
temperature $(T_L+T_R)/2$ is selected.
\et

\bpr
We start with the following lemma
\bl
For all $\xi\in\Omega^d_N$ a (dual) particle configuration,
there exists
$c=c(\xi)>0$, $a=a(\xi)>0$ such that
for all $\lambda >0$, and for all $t>0$
\[
 \left| \int  \E_x D(\xi,x_t) \ \nu^\lambda_{T_L,T_R} (dx) -  \int D(\xi,x) \ \mu^\lambda_{T_L,T_R} (dx) \right|
\leq ce^{-\lambda a t}  \]
\el
\bpr
Using duality between $BMP_\lambda$ and $SIP_\lambda$, and \eqref{statmeas}
\begin{eqnarray*}
&& \left| \int  \E_x D(\xi,x_t) \ \nu^\lambda_{T_L,T_R} (dx) -  \int D(\xi,x) \ \mu^\lambda_{T_L,T_R} (dx) \right|
\\
&=&\left| \E_{\xi} \left(\prod_{i=1}^N (T_{i}^{(\lambda)})^{\xi_{i}(t)} T_{L}^{\xi_{0}(t)}T_{R}^{\xi_{N+1}(t)}\right) - \E_{\xi}\left( T_{L}^{\xi_{0}(\infty)}T_{R}^{\xi_{N+1}(\infty)}\right) \right|
\\
& \leq& C(\xi) \pee_{\xi}^d (\xi(t) \neq \xi(\infty))
\\
& \leq& C(\xi) \pee_{\xi}^d (\text{ there exist particles that are not absorbed at time t})
\\
& \leq & C(\xi) e^{-a\lambda t}
\end{eqnarray*}
In order to see the last inequality, we remark that for a
particle at positions $1,N$, the probability to be
absorbed at the next step is of order $\lambda$, as the
maximal rate to move to the other (non-absorbing) neighbor
is at most $2(|\xi|+1)$.
\epr

Proof of Theorem \ref{lambdathm}: using Lemma 6.1, and duality between $SIP_\lambda$ and $BMP_\lambda$,
we have
\begin{eqnarray*}
 \left|\int  \E_x D(\xi,x_t) \ \nu^\lambda_{T_L,T_R} (dx) -  \int D(\xi,x) \ \nu^\lambda_{T_L,T_R} (dx)  \right|
&=&\left| \int_0^t L_d \psi (\xi_s) ds\right|\\
&\leq & C(|\xi|) \lambda^2 t
\end{eqnarray*}
Combining with Lemma 6.2 we have
\begin{equation}
 \int  D(\xi,x_t) \ \nu^\lambda_{T_L,T_R} (dx) -  \int D(\xi,x) \ \mu^\lambda_{T_L,T_R} (dx)  
\leq  C(\xi) \left( \lambda^2 t + e^{-a\lambda t} \right)
\end{equation}
Now optimize w.r.t. $t$ by choosing $t=(1/a\lambda) \log (a/\lambda)$
\epr
%From the proof of the theorem, we obtain that for all $\epsi>0$
%\be\label{papegaai}
%\left|\int D(\xi,x)\nu^\lambda_{T_L,T_R}-\int D(\xi,x)\nu^\lambda_{T_L,T_R}\right|
%\leq C_\epsi(\xi) \lambda^{1-\epsi}
%\ee
%we will complement this in the next section with the fact that the two-point
%correlation function is at least of order $\lambda$. 
\section{The two point correlation functions in the limit $\lambda\rightarrow0$}
In this section we prove that for the two-point correlation function in the non-equilibrium steady state, the 
deviation from local equilibrium is of order $\lambda$, which
strengthens \eqref{papegaai} for $\xi=\delta_i+\delta_j$ (i.e., we get rid
of the $\log(1/\lambda)$-factor). In the appendix
we give explicit expressions for the two-point function of
some finite systems, and show in particular that it is not 
multilinear for $\lambda\not=1$.

Define for $i,j \in \{1,\ldots,N\}$
\[
{\bf Y}_{ij}=\int(x_{i}^{2}x_{j}^{2}) \mu_{T_{L}T_{R}}^{(\lambda)}(dx)\]
and additionally ${\bf Y}_{0i}=T_{L}T_{i}$, ${\bf Y}_{i,N+1}=T_{i}T_{R}$.

Denote by ${\bf T}$ the matrix with elements
${\bf T}_{ij}=T_{i}T_{j}$ if $i\neq j$ and ${\bf T}_{ij}=3T_{i}^{2}$ if $i=j$ 
where $T_i$ is the temperature profile of Theorem \ref{tempthm}
%, expanded up to first order
%in $\lambda$, i.e.,
%\[
%T_{i}=\frac{T_{L}+T_{R}}{2}+\frac{\lambda(N+1)(T_{L}-T_{R})}{4}+\frac{\lambda(T_{R}-T_{L})}{2}i\]

\bt\label{twopointthm}
There exists $C>0$ such that
for all $i,j\in \{1,\ldots,N\}$ we
have
\be\label{kanarie}
|{\bf Y}_{ij}- {\bf T}_{ij}| \leq C \lambda
\ee
\et
\bpr
From the stationarity of $\mu^\lambda_{T_L,T_R}$ we find that
${\bf Y}$ satisfies the following system of linear equations for $k,l \in \{1,\ldots,N\}$
\begin{eqnarray}\label{twopointeq}
0&=&(-4Y_{kl}+Y_{k-1l}+Y_{k+1l}+Y_{kl-1}+Y_{kl+1})
\nonumber\\
&+& 4Y_{kk+1}\delta_{kl}+4Y_{k-1k}\delta_{kl}-4Y_{k-1k}\delta_{k,l+1}-4Y_{kk+1}\delta_{k,l-1}
\nonumber\\
&+&\lambda(T_{L}T_{l}-Y_{1l})\delta_{1k}+\lambda(T_{L}T_{k}-Y_{1k})\delta_{1l}
\nonumber\\
&+&\lambda(T_{R}T_{l}-Y_{Nl})\delta_{Nk}+\lambda(T_{R}T_{k}-Y_{Nk})\delta_{Nl}
\end{eqnarray}
which has the form
\[
{\bf M.Y} ={\bf D}\]
By explicit computation we obtain
\be\label{bozen}
{\bf X:=M.T-D} =O(\lambda^{2})  
\ee
%where for a $A$ is a matrix which does not vanish in the limit $\lambda\to 0$.
From this we will now derive that
\be\label{yyy}
{\bf Y=T}+O(\lambda).
\ee
%which strengthens the results of theorem \ref{} for the two-point function.
%\[
%\lim_{\lambda\rightarrow0}\; Y^{(\lambda)}=T\]

Put
\[
{\bf \|Y-T\|=\|M^{-1}M(Y-T)\|=\|M^{-1}X\|}\]

We will show that
\be\label{abcd}
\| {\bf M^{-1}X }\|^{2}\leq\frac{c}{\lambda^{2}}\|{\bf X } \|^{2}
\ee
which combined with \eqref{bozen} gives the desired result \eqref{yyy}.

To obtain \eqref{abcd} consider
\begin{eqnarray*}
<{\bf M} ^{-1}{\bf X} , {\bf M} ^{-1}{\bf X}>&=&<{\bf X},({\bf M}^{-1})^{T}{\bf M}^{-1}{\bf X}>
\\
&=&<{\bf X}, {\bf A}^{-1}{\bf X}>
\end{eqnarray*}
with ${\bf A}:= {\bf MM} ^{T}$
Using the spectral decomposition of ${\bf A}$, we get
\begin{eqnarray*}
<{\bf X}, {\bf A}^{-1}{\bf X}>&=&\sum_{i}\frac{1}{\lambda_{i}^{({\bf A})}}<{\bf X},{\bf e}_{i}><{\bf e}_{i},{\bf X}>
\\
&\leq&\frac{1}{\min_{i}(\lambda_{i}^{({\bf A})})}||{\bf X}||^{2}
\end{eqnarray*}
where $\lambda_{i}$ are the eigenvalues of ${\bf A}$ with the corresponding eigenvectors ${\bf e}_{i}$.
So it suffices now to see that
\[
\min_{i}(\lambda_{i}^{({\bf A})})\geq c\lambda^{2}\]
We have
\[
\min_{i}(\lambda_{i}^{({\bf A})})=\inf_{||{\bf X}||=1}<{\bf X,AX}>\]
The matrix ${\bf M}$ has the form ${\bf M= K}+\lambda {\bf S}$
and hence
\[
<{\bf X,AX}>=<({\bf K}^{T}+\lambda {\bf S}^{T}){\bf X},({\bf K}^{T}+\lambda {\bf S}^{T}){\bf X}>\]
Therefore
\[
\frac{<{\bf X,AX}>}{\lambda^{2}}=\frac{\lambda^{2}||{\bf S}^{T}{\bf X}||^{2}+2\lambda<{\bf S}^{T}{\bf X},{\bf K}^{T}{\bf X}>+||{\bf K}^{T}{\bf X}||^{2}}{\lambda^{2}}\]

and so we obtain
\[
\liminf_{\lambda\rightarrow0}\;\frac{\min_{i}(\lambda_{i}^{({\bf A})})}{\lambda^{2}}>0\]
Indeed,
since ${\bf M}\equiv {\bf K}+\lambda {\bf S}$ is not singular, either ${\bf K}$ or ${\bf S}$
must not be singular, therefore $||{\bf S}^{T} {\bf X}||^{2}$ and $||{\bf K}^{T}{\bf X}||^{2}$ cannot be both zero.
\epr
\br
It follows from the correlation inequalities derived
in \cite{grv} that ${\bf Y}_{ij}\geq {\bf T}_{ij}$. Indeed, ${\bf T}_{ij}$ would
be the correlation function if the dual walkers were walking independently, however, two dual
walkers interact
by inclusion (attraction), and this leads to a positive covariance.
\er

\section{Acknowledgment}
We would like to thank Christian Giardina for usefull discussions.

\section{Appendix}
Here we derive explicit expressions for the two point correlation function
for systems with three and four sites.
We start from the equations \eqref{twopointeq}.

Since $Y_{kl}$ is symmetric in $k$ and $l$ it suffices to consider $k\leq l$. the different cases are as follows;
\begin{enumerate}
\item $\bf 1<k=l<N$; $(-2Y_{kk}+3Y_{kk-1}+3Y_{kk+1})=0$
\item $\bf 1<k=l-1<N-1$; $(-8Y_{kk+1}+Y_{k-1k+1}+Y_{k+1k+1}+Y_{kk}+Y_{kk+2})=0$
\item $\bf 1<k<l+1<N+1$; $(-4Y_{kl}+Y_{k-1l}+Y_{k+1l}+Y_{kl-1}+Y_{kl+1})=0$
\item $\bf k=l=1$; $(-2Y_{11}+3Y_{10}+3Y_{12})+\lambda(T_{L}T_{1}-Y_{11})+\lambda(T_{L}T_{1}-Y_{11})=0$
\item $\bf k=l=N$; $(-2Y_{NN}+3Y_{NN-1}+3Y_{NN+1})+\lambda(T_{R}T_{N}-Y_{NN})+\lambda(T_{R}T_{N}-Y_{NN})=0$
\item $\bf 1=k=l-1$; $(-8Y_{12}+Y_{02}+Y_{22}+Y_{11}+Y_{13})+\lambda(T_{L}T_{2}-Y_{12})=0$
\item $\bf k=l-1=N-1$; $(-8Y_{N-1N}+Y_{N-2N}+Y_{NN}+Y_{N-1N-1}+Y_{N-1N+1})+\lambda(T_{R}T_{N-1}-Y_{NN-1})=0$
\item $\bf 1=k<l+1<N+1$; $(-4Y_{1l}+Y_{0l}+Y_{2l}+Y_{1l-1}+Y_{1l+1})+\lambda(T_{L}T_{1}-Y_{1l})=0$
\item $\bf 1=k<l+1=N+1$; $(-4Y_{1N}+Y_{0N}+Y_{2N}+Y_{1N-1}+Y_{1N+1})+\lambda(T_{L}T_{1}-Y_{1N})+\lambda(T_{R}T_{1}-Y_{1N})=0$
\item $\bf 1<k<l+1=N+1$; $(-4Y_{kN}+Y_{k-1N}+Y_{k+1N}+Y_{kN-1}+Y_{kN+1})+\lambda(T_{R}T_{k}-Y_{kN})=0$
\end{enumerate}

\subsection{3 Sites System}

The equations for the two-point correlation function are of the form ${\bf M.Y=D}$
where

${\bf Y}=\left(\begin{array}{c}
Y_{11}\\
Y_{12}\\
Y_{13}\\
Y_{22}\\
Y_{23}\\
Y_{33}\end{array}\right)$ and ${\bf D}=\left(\begin{array}{c}
-\lambda T_{L}T_{3}-\lambda T_{R}T_{1}\\
-3\lambda T_{R}T_{3}\\
-\lambda T_{R}T_{2}\\
-3\lambda T_{L}T_{1}\\
-\lambda T_{L}T_{2}\\
0\end{array}\right)$

and the matrix ${\bf M}$ can be read from the previous equations as;

\[
{\bf M}=\left(\begin{array}{cccccc}
0 & 1 & -2(1+\lambda) & 0 & 1 & 0\\
0 & 0 & 0 & 0 & 3 & -(1+\lambda)\\
0 & 0 & 1 & 1 & -(7+\lambda) & 1\\
-(1+\lambda) & 3 & 0 & 0 & 0 & 0\\
1 & -(7+\lambda) & 1 & 1 & 0 & 0\\
0 & 3 & 0 & -2 & 3 & 0\end{array}\right)\]

The explicit solution is 
via inversion of ${\bf M}$.
The result for ${\bf Y}$ and the correlation functions ${\bf C}_{ij}= {\bf Y}_{ij}-T_i T_j (1+2\delta_{ij})$ then reads as follows:

\[
{\bf Y}_{11}=\frac{3\left(T_{R}^{2}(5+3\lambda)+2T_{L}T_{R}\left(5+9\lambda+2\lambda^{2}\right)+T_{L}^{2}\left(5+23\lambda+24\lambda^{2}+4\lambda^{3}\right)\right)}{4(1+\lambda)^{2}(5+\lambda)}\]
\[
{\bf Y}_{12}=\frac{T_{R}^{2}(5+3\lambda)+2T_{L}T_{R}\left(5+4\lambda+\lambda^{2}\right)+T_{L}^{2}\left(5+13\lambda+2\lambda^{2}\right)}{4\left(5+6\lambda+\lambda^{2}\right)}\]
\[
{\bf Y}_{13}=\frac{T_{L}^{2}\left(5+13\lambda+2\lambda^{2}\right)+T_{R}^{2}\left(5+13\lambda+2\lambda^{2}\right)+2T_{L}T_{R}\left(5+9\lambda+12\lambda^{2}+2\lambda^{3}\right)}{4(1+\lambda)^{2}(5+\lambda)}\]
\[
{\bf Y}_{22}=\frac{3\left(2T_{L}T_{R}\left(5+4\lambda+\lambda^{2}\right)+T_{L}^{2}\left(5+8\lambda+\lambda^{2}\right)+T_{R}^{2}\left(5+8\lambda+\lambda^{2}\right)\right)}{4\left(5+6\lambda+\lambda^{2}\right)}\]
\[
{\bf Y}_{23}=\frac{T_{L}^{2}(5+3\lambda)+2T_{L}T_{R}\left(5+4\lambda+\lambda^{2}\right)+T_{R}^{2}\left(5+13\lambda+2\lambda^{2}\right)}{4\left(5+6\lambda+\lambda^{2}\right)}\]
\[
{\bf Y}_{33}=\frac{3\left(T_{L}^{2}(5+3\lambda)+2T_{L}T_{R}\left(5+9\lambda+2\lambda^{2}\right)+T_{R}^{2}\left(5+23\lambda+24\lambda^{2}+4\lambda^{3}\right)\right)}{4(1+\lambda)^{2}(5+\lambda)}\]
and
\[
{\bf C}_{11}=\frac{3(T_{L}-T_{R})^{2}\lambda}{2(1+\lambda)^{2}(5+\lambda)},{\bf C}_{12}=\frac{(T_{L}-T_{R})^{2}\lambda}{2\left(5+6\lambda+\lambda^{2}\right)}\]

\[
{\bf C}_{13}=\frac{(T_{L}-T_{R})^{2}\lambda}{2(1+\lambda)^{2}(5+\lambda)},{\bf C}_{22}=\frac{3(T_{L}-T_{R})^{2}\lambda}{2\left(5+6\lambda+\lambda^{2}\right)}\]

\[
{\bf C}_{23}=\frac{(T_{L}-T_{R})^{2}\lambda}{2\left(5+6\lambda+\lambda^{2}\right)},{\bf C}_{33}=\frac{3(T_{L}-T_{R})^{2}\lambda}{2(1+\lambda)^{2}(5+\lambda)}\]

We see that for all $k,l$
\[
{\bf C}_{kl}\propto\lambda(T_{L}-T_{R})^{2}\]
and also ${\bf C}_{kl}\geq 0$.

One might be interested to see if the bi-linear ansatz introduced
in \cite{gkr} for the special case $\lambda=1$ is also valid here, i.e.
\begin{eqnarray}\label{bili}
{\bf Y}_{ij}&=& a+bi+cj+dij\nonumber\\
{\bf Y}_{ii}&=& A+Bi+Di^{2}
\end{eqnarray}

with the boundary conditions 
${\bf Y}_{0i}=T_{L}T_{i}$, ${\bf Y}_{i,N+1}=T_{i}T_{R}$.

However, to check the validity of the ansatz we must calculate the
correlation functions for a 4 sites system, since in 3 sites systems 
we have only 6 correlation functions which are less than the 7 constants
of the ansatz.

\subsection{4 Sites System}

Similar to the calculation for the 3 site system, we have ${\bf M.Y=D}$
where

${\bf Y}=\left(\begin{array}{c}
Y_{11}\\
Y_{12}\\
Y_{13}\\
Y_{14}\\
Y_{22}\\
Y_{23}\\
Y_{24}\\
Y_{33}\\
Y_{34}\\
Y_{44}\end{array}\right)$ and ${\bf D}=\left(\begin{array}{c}
0\\
0\\
0\\
\text{ }-\lambda T_{L}T_{3}\\
-\lambda T_{L}T_{2}\\
-3\lambda T_{L}T_{1}\\
-\lambda T_{2}T_{R}\text{ }\\
-\lambda T_{R}T_{3}\\
-3\lambda T_{4}T_{R}\\
-\lambda T_{L}T_{4}-\lambda T_{1}T_{R}\end{array}\right)$\\
and where the matrix ${\bf M}$ is given by
\\
$
\left(\begin{array}{cccccccccc}
0 & 0 & 1 & 0 & 1 & -8 & 1 & 1 & 0 & 0\\
0 & 3 & 0 & 0 & -2 & 3 & 0 & 0 & 0 & 0\\
0 & 0 & 0 & 0 & 0 & 3 & 0 & -2 & 3 & 0\\
0 & 1 & -3-\lambda & 1 & 0 & 1 & 0 & 0 & 0 & 0\\
1 & -7-\lambda & 1 & 0 & 1 & 0 & 0 & 0 & 0 & 0\\
-1-\lambda & 3 & 0 & 0 & 0 & 0 & 0 & 0 & 0 & 0\\
0 & 0 & 0 & 1 & 0 & 1 & -3-\lambda & 0 & 1 & 0\\
0 & 0 & 0 & 0 & 0 & 0 & 1 & 1 & -7-\lambda & 1\\
0 & 0 & 0 & 0 & 0 & 0 & 0 & 0 & 3 & -1-\lambda\\
0 & 0 & 1 & -2-2\lambda & 0 & 0 & 1 & 0 & 0 & 0\end{array}\right)
$
The solution for ${\bf Y}$ is
\[
\begin{array}{c}
{\bf Y}_{11}=\frac{6T_{R}^{2}(12+\lambda(14+3\lambda))+6T_{L}T_{R}(24+\lambda(76+5\lambda(8+\lambda)))+T_{L}^{2}(72+3\lambda(172+3\lambda(106+\lambda(46+5\lambda))))}{(6+\lambda)(2+3\lambda)(8+\lambda(16+5\lambda))}\\
{\bf Y}_{12}=\frac{2T_{R}^{2}(1+\lambda)(12+\lambda(14+3\lambda))+T_{L}T_{R}(48+\lambda(152+\lambda(128+\lambda(44+5\lambda))))+2T_{L}^{2}(12+\lambda(74+\lambda(121+\lambda(49+5\lambda))))}{(6+\lambda)(2+3\lambda)(8+\lambda(16+5\lambda))}\\
{\bf Y}_{13}=\frac{T_{L}^{2}(1+\lambda)(24+5\lambda(4+\lambda)(5+\lambda))+T_{R}^{2}(24+\lambda(76+\lambda(41+4\lambda)))+2T_{L}T_{R}(24+\lambda(76+\lambda(109+\lambda(47+5\lambda))))}{(6+\lambda)(2+3\lambda)(8+\lambda(16+5\lambda))}\\
{\bf Y}_{14}=\frac{T_{L}^{2}(24+5\lambda(4+\lambda)(5+\lambda))+T_{R}^{2}(24+5\lambda(4+\lambda)(5+\lambda))+T_{L}T_{R}(48+\lambda(152+\lambda(314+3\lambda(46+5\lambda))))}{(6+\lambda)(2+3\lambda)(8+\lambda(16+5\lambda))}\\
{\bf Y}_{22}=\frac{3\left(2T_{L}T_{R}(24+\lambda(76+\lambda(73+3\lambda(9+\lambda))))+T_{L}^{2}(24+\lambda(124+\lambda(181+7\lambda(10+\lambda))))+T_{R}^{2}(24+\lambda(76+\lambda(77+2\lambda(12+\lambda))))\right)}{(6+\lambda)(2+3\lambda)(8+\lambda(16+5\lambda))}\\
{\bf Y}_{23}=\frac{2T_{L}^{2}(12+\lambda(5+2\lambda)(10+\lambda(8+\lambda)))+2T_{R}^{2}(12+\lambda(5+2\lambda)(10+\lambda(8+\lambda)))+T_{L}T_{R}(2+\lambda)(24+\lambda(64+\lambda(50+7\lambda)))}{(6+\lambda)(2+3\lambda)(8+\lambda(16+5\lambda))}\\
{\bf Y}_{24}=\frac{T_{R}^{2}(1+\lambda)(24+5\lambda(4+\lambda)(5+\lambda))+T_{L}^{2}(24+\lambda(76+\lambda(41+4\lambda)))+2T_{L}T_{R}(24+\lambda(76+\lambda(109+\lambda(47+5\lambda))))}{(6+\lambda)(2+3\lambda)(8+\lambda(16+5\lambda))}\\
{\bf Y}_{33}=\frac{3\left(2T_{L}T_{R}(24+\lambda(76+\lambda(73+3\lambda(9+\lambda))))+T_{R}^{2}(24+\lambda(124+\lambda(181+7\lambda(10+\lambda))))+T_{L}^{2}(24+\lambda(76+\lambda(77+2\lambda(12+\lambda))))\right)}{(6+\lambda)(2+3\lambda)(8+\lambda(16+5\lambda))}\\
{\bf Y}_{34}=\frac{2T_{L}^{2}(1+\lambda)(12+\lambda(14+3\lambda))+T_{L}T_{R}(48+\lambda(152+\lambda(128+\lambda(44+5\lambda))))+2T_{R}^{2}(12+\lambda(74+\lambda(121+\lambda(49+5\lambda))))}{(6+\lambda)(2+3\lambda)(8+\lambda(16+5\lambda))}\\
{\bf Y}_{44}=\frac{6T_{L}^{2}(12+\lambda(14+3\lambda))+6T_{L}T_{R}(24+\lambda(76+5\lambda(8+\lambda)))+3T_{R}^{2}(24+\lambda(172+3\lambda(106+\lambda(46+5\lambda))))}{(6+\lambda)(2+3\lambda)(8+\lambda(16+5\lambda))}\end{array}\]
and the corresponding correlation functions are

\[
\begin{array}{c}
{\bf C}_{11}=\frac{3(T_{L}-T_{R})^{2}\lambda(24+\lambda(50+13\lambda))}{(6+\lambda)(2+3\lambda)^{2}(8+\lambda(16+5\lambda))},
{\bf C}_{12}=\frac{(T_{L}-T_{R})^{2}\lambda(1+\lambda)(24+\lambda(50+13\lambda))}{(6+\lambda)(2+3\lambda)^{2}(8+\lambda(16+5\lambda))}\\
{\bf C}_{13}=\frac{2(T_{L}-T_{R})^{2}\lambda(1+\lambda)(12+\lambda(16+\lambda))}{(6+\lambda)(2+3\lambda)^{2}(8+\lambda(16+5\lambda))},
{\bf C}_{14}=\frac{2(T_{L}-T_{R})^{2}\lambda(12+\lambda(16+\lambda))}{(6+\lambda)(2+3\lambda)^{2}(8+\lambda(16+5\lambda))}\\
{\bf C}_{22}=\frac{3(T_{L}-T_{R})^{2}\lambda(2+\lambda(4+\lambda))(12+\lambda(16+\lambda))}{(6+\lambda)(2+3\lambda)^{2}(8+\lambda(16+5\lambda))},
{\bf C}_{23}=\frac{(T_{L}-T_{R})^{2}\lambda(24+\lambda(86+\lambda(93+\lambda(27+2\lambda))))}{(6+\lambda)(2+3\lambda)^{2}(8+\lambda(16+5\lambda))}\\
{\bf C}_{24}=\frac{2(T_{L}-T_{R})^{2}\lambda(1+\lambda)(12+\lambda(16+\lambda))}{(6+\lambda)(2+3\lambda)^{2}(8+\lambda(16+5\lambda))},
{\bf C}_{33}=\frac{3(T_{L}-T_{R})^{2}\lambda(2+\lambda(4+\lambda))(12+\lambda(16+\lambda))}{(6+\lambda)(2+3\lambda)^{2}(8+\lambda(16+5\lambda))}\\
{\bf C}_{34}=\frac{(T_{L}-T_{R})^{2}\lambda(1+\lambda)(24+\lambda(50+13\lambda))}{(6+\lambda)(2+3\lambda)^{2}(8+\lambda(16+5\lambda))},
{\bf C}_{44}=\frac{3(T_{L}-T_{R})^{2}\lambda(24+\lambda(50+13\lambda))}{(6+\lambda)(2+3\lambda)^{2}(8+\lambda(16+5\lambda))}\end{array}\]
We see once more that  for all $k,l$
\[
{\bf C}_{kl}\propto\lambda(T_{L}-T_{R})^{2}\]
and ${\bf C}_{kl} \geq 0$.

Now we can directly check the validity of the bi-linear ansatz. 
Direct calculation shows that the diagonal part of
the ansatz, i.e., ${\bf Y}_{ii}=A+Bi+Di^{2}$ is valid, but the non-diagonal
part ${\bf Y}_{ij}=a+bi+cj+dij$ is not.

If we determine the coeficients $a,b,c,d$ by fitting the bilinear ansatz to ${\bf Y}_{12},{\bf Y}_{13},{\bf Y}_{23},{\bf Y}_{34}$, then
we obtain
 \[
{\bf Y}_{14}-\left(a+b+4c+4d\right)=\frac{3(T_{L}-T_{R})^{2}(-1+\lambda)\lambda^{2}}{(6+\lambda)(2+3\lambda)(8+\lambda(16+5\lambda))}.
\]
which shows that the bilinear form can not hold for $\lambda\not\in  \{0,1\}$. Remark that also when $\lambda\to\infty$ the
deviation from the multilinear form vanishes, which is consistent with the intuition that this limit
is the same as having $\lambda=1$ in a smaller system obtained by removing the sites $1,N$.

\end{document}